\documentclass[aps,prb,preprint,showpacs]{revtex4-1}
\usepackage[utf8]{inputenc}
\usepackage{amsmath,amssymb,amsfonts,amsthm,array,dcolumn,nicefrac,units}
\usepackage{graphicx}

\newcommand{\OO}{\hat{\mathcal{O}}}
\newcommand{\HH}{\mathcal{H}}
\newcommand{\Ssm}[2]{\langle\hat{\mathbf{S}}\cdot\hat{\mathbf{s}}\rangle_{#1,#2}}
\newcommand{\unity}{\hat{\mathbf{1}}}

\begin{document}

\title{Analysis of the inelastic contribution to atomic spin excitation spectroscopy}
\author{Bruno Chilian}
\affiliation{Institute of Applied Physics, Hamburg University, Jungiusstrasse 11, D-20355 Hamburg, Germany}
\author{Alexander A. Khajetoorians}
\affiliation{Institute of Applied Physics, Hamburg University, Jungiusstrasse 11, D-20355 Hamburg, Germany}
\author{Jens Wiebe}
\email[corresponding author.\\Email adress: ]{jwiebe@physnet.uni-hamburg.de}
\affiliation{Institute of Applied Physics, Hamburg University, Jungiusstrasse 11, D-20355 Hamburg, Germany}
\author{Roland Wiesendanger}
\affiliation{Institute of Applied Physics, Hamburg University, Jungiusstrasse 11, D-20355 Hamburg, Germany}


\begin{abstract}
We investigate the dependence of the inelastic spin excitation contribution to scanning tunneling spectroscopy recorded above Fe adsorbates on the InSb(110) surface at subkelvin temperatures, on bulk doping and tunnel junction resistance. To explain our observations, we show how the inelastic contribution depends on the parameters describing the excitation mechanism and tunnel conditions, in the framework of a recently developed model. We conclude that in this particular system of an adsorbate which is strongly relaxed into the substrate, the tip-sample distance dependent by-tunneling has to be taken into account in order to explain the observed variations in the inelastic tunnel spectra. 
\end{abstract}

\maketitle

In the quest for nanometer scale spintronic devices, the scanning tunneling microscope (STM) is a valuable tool due to its unrivaled  ability to locally probe and manipulate the magnetic structure of the sample system with atomic resolution. 
Most notably, inelastic scanning tunneling spectroscopy~\cite{HeinrichScience2004,HirjibehedinScience2007,PhysRevLett.101.197208,KawaiPRL2010} (ISTS) and spin-polarized scanning tunneling spectroscopy~\cite{MZW2008} (SP-STS) experiments have contributed decisively to the understanding of single magnetic adsorbates on surfaces and their interactions~\cite{HirjibehedinScience2006,ZhouWiebeLounis2010}. 

Understanding the mechanism governing ISTS experiments has been the purpose of numerous theoretical advances \cite{LorentePhysRevLett.103.176601,PhysRevB.81.165423,Fernandez_Rossier_PhysRevLett.102.256802,Fransson_Nanolett2009,Persson_PhysRevLett.103.050801,NewJPhys.12.125021}, which relate the experimentally accessible excitation spectra to the unobservable quantum processes that occur during tunneling. The descriptions range from a generalized Anderson model~\cite{PhysRev.149.491,PhysRev.154.633} in a co-tunneling picture~\cite{NewJPhys.12.125021} commonly used in transport theory~\cite{PhysRevLett.86.878} to approaches that treat the excitation event in a sudden approximation~\cite{LorentePhysRevLett.103.176601,PhysRevB.81.165423}, with a formalism similar to scattering theory. All of these models accurately reproduce the measured intensities of a given measurement, opening up the possibility of probing the hidden spin-excitation mechanism by comparison of model and experiment.

In this work, we analyze ISTS and magnetic field dependent spin-split Landau level spectroscopy of single Fe adsorbates on the InSb(110) surface containing a two-dimensional electron system (2DES)~\cite{KhajetooriansNature2010} for differently doped substrates and for different junction resistances. 
We observe a significant variation in the inelastic contribution to our spectra with the doping concentration and junction resistance. To explain our observations, we utilize the model which treats the excitation events in a sudden approximation, where the formation and decay of an intermediate total spin state of tunneling electron and adsorbate governs the excitation probabilities~\cite{LorentePhysRevLett.103.176601}. We show for an arbitrary spin, how the inelastic signal percentage $I_{\rm inel}$ and the 2DES Landau level asymmetry $A_{\rm LL}$ depend on the model parameters that describe the excitation mechanism and tunnel conditions. By combining the measured $I_{\rm inel}$ and $A_{\rm LL}$, it is then in principle possible to extract the model parameters. We discuss the various physical effects that determine $I_{\rm inel}$ and $A_{\rm LL}$ and conclude that in our case, distance-dependent by-tunneling is most likely the origin of the variations in $I_{\rm inel}$.

All measurements were performed in an ultra-high vacuum (UHV) STM facility with a magnetic field of up to $\unit[12]{T}$ perpendicular to the sample surface (out-of plane) operated at a temperature of $\unit[300]{mK}$~\cite{WWM2004}. 
We used electrochemically etched W tips, which were flashed in UHV to a temperature of approx. $\unit[2000]{K}$. The $n$-doped InSb crystals with doping concentrations in the range of $\unit[0.5 - 2\cdot 10^{16}]{cm^{-3}}$ were cleaved in UHV to expose the clean, atomically flat (110) surface~\cite{HSW2008}, onto which single Fe atoms were deposited at low temperature ($T<\unit[25]{K}$)~\cite{KhajetooriansNature2010}. To obtain the $\text{d}I/\text{d}V$-spectra, the feedback circuit was switched off at a stabilization current $I_{\rm stab}$ and sample voltage $V_{\rm stab}$ of $\unit[4 - 20]{mV}$ for the ISTS and $\unit[100]{mV}$ for the 2DES/Landau level spectra.  The sample voltage $V$ was ramped while recording the differential conductance signal via a lock-in technique  with a modulation voltage $V_{\rm mod}$ (rms value) added to $V$ ($f = \unit[828]{Hz}$).

\begin{figure}
\includegraphics[width=8.6cm]{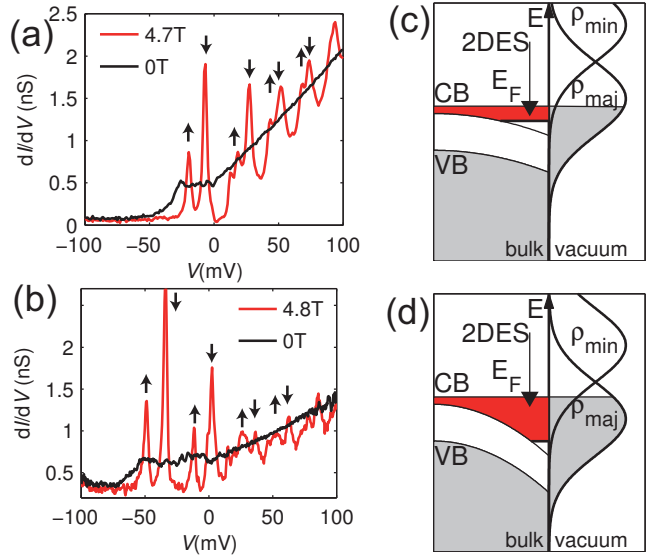}
\caption{
(a) Typical 2DES and LL spectra for the high doped sample, exhibiting weak band bending. The 2DES (black) and the LL (red) spectra, were taken in zero magnetic field, and in a magnetic field $B_{y}$ perpendicular to the surface, respectively. ($I_{\rm stab}=\unit[200]{pA}$, $V_{\rm stab}=\unit[100]{mV}$, $V_{\rm mod}=\unit[1]{mV}$). (b) The same for the low doped sample exhibiting strong band bending, ($I_{\rm stab}^{2DES}=\unit[200]{pA}$, $I_{\rm stab}^{LL}=\unit[150]{pA}$, $V_{\rm stab}=\unit[100]{mV}$, $V_{\rm mod}=\unit[0.5]{mV}$). The 2DES spectrum was multiplied by $3/4$ to compensate for the different $I_{\rm stab}$. The spin states of the spin-split LLs are indicated by arrows.
(c)/(d) Schematic representation of the band bending for high/low doping. The conduction band (CB) and valence band (VB) edges are energetically lower at the surface (vertical line) than in the bulk, inducing a surface 2DES. An artificial distribution of the majority and minority vacuum LDOS of the Fe adsorbate is sketched. 
\label{fig:1}}\end{figure}

A 2DES is induced at the surface by the adsorbate-induced downwards band bending~\cite{HSW2008}. We varied the band bending by choosing samples with two different doping levels, one higher doped with a carrier concentration (at 77K) of $\unit[2\cdot 10^{16}]{cm^{-3}}$ and one lower doped with $\unit[6.5\cdot 10^{15}]{cm^{-3}}$. The spectroscopic signature of the 2DES is a step in the differential conductance with an onset in the bulk band gap [$\unit[-240]{mV}$ to $\unit[0]{mV}$] which indicates the energy of the lowest subband $E_{1}$ of the 2DES, as is shown in Fig.~\ref{fig:1}(a,b). From the difference in the lowest subband energy, which is $E_1 \approx \unit[-30]{meV}$ for the high doped, and $E_1 \approx \unit[-55]{meV}$ for the low doped sample, the difference in band bending $\Delta_{\rm BB}\approx\unit[25]{meV}$ between the two samples was estimated. Fig.~\ref{fig:1} (c)/(d) shows a schematic sketch of the tunneling conditions for weak/strong downwards band bending. 
Schematic sketches of the minority ($\rho_{\rm min}$) and majority ($\rho_{\rm maj}$) local vacuum densities of states (LDOS) above the Fe adsorbate, i.e. at the location of the tip, are included. Figures 1(a,b) also show $\text{d}I/\text{d}V$ curves measured in a magnetic field of $B\approx\unit[5]{T}$ revealing spin-split Landau levels (LL) as discussed later.

\begin{figure}
\includegraphics[width=8.6cm]{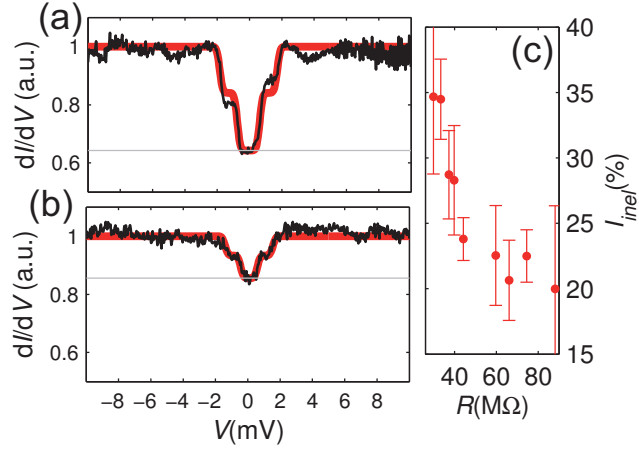}
\caption{
(a)/(b) Typical ISTS spectra of isolated Fe adsorbates taken at $B_{y} = 0$ (T = \unit[0.3]{K}). The spectra were normalized by division by a spectrum taken at a nearby substrate location. Broad red curves represent the best fit to a thermally broadened symmetric double-step function and gray horizontal lines mark the inelastic contribution $I_{\rm inel}$ to the total differential conductance (see text). (a) Higher doped sample ($I_{\rm stab}=\unit[600]{pA}$, $V_{\rm stab}=\unit[20]{mV}$, $V_{\rm mod}=\unit[40]{\mu V}$). (b) Lower doped sample ($I_{\rm stab}=\unit[200]{pA}$, $V_{\rm stab}=\unit[10]{mV}$, $V_{\rm mod}=\unit[40]{\mu V})$.
(c) $I_{\rm inel}$ as a function of junction resistance $R$ on the high doped sample. The error bars were computed from a $\unit[90]{\%}$-confidence interval to the fit parameters.
\label{fig:2}}\end{figure}

The results obtained from the ISTS measurements for the high and low doped samples are shown in Fig.~\ref{fig:2} (a,b). For both samples, the spectra reveal two steps above and below the Fermi energy $E_{\rm F}$ ($V = \unit[0]{V}$), which are symmetric in energy with respect to $E_{\rm F}$. These steps are due to the inelastic excitations of the adsorbate spin by the tunneling electrons~\cite{KhajetooriansNature2010}. However, the intensity of the steps strongly differs between the two samples, i.e. the inelastic contribution $I_{\rm inel}$ to the spectra differs. The spectra were normalized by dividing them by a spectrum taken on a nearby substrate location. $I_{\rm inel}$ was extracted from the normalized spectra by fitting a thermally broadened symmetric double step function and then taking the difference of the function above ($V>\unit[5]{mV}$) and below ($V=\unit[0]{mV}$) all excitations [gray horizontal line in Fig.~\ref{fig:2} (a,b)]. For the higher doped sample, measured with a junction resistance $R=\frac{U_{\rm stab}}{I_{\rm stab}}=\unit[30]{M\Omega}$, we obtained $I_{\rm inel}=\unit[31]{\%}\pm\unit[4]{\%}$ (standard deviation) and for the lower doped sample, measured with $R=\unit[50]{M\Omega}$, $I_{\rm inel}=\unit[16]{\%}\pm\unit[4]{\%}$, when analyzing a total of 24 adsorbates. 
To be able to separate the influence of doping and junction resistance, we recorded $I_{\rm inel}$ as a function of $R$ on the higher doped sample (Fig.~\ref{fig:2}(c)). We observe that at a junction resistance $R=\unit[50]{M\Omega}$ on the higher doped sample, $I_{\rm inel}\approx\unit[20 - 25]{\%}$, which is only slightly larger than the $I_{\rm inel}$ measured on the lower doped sample at the same junction resistance. Considering the error margins, this implies that changing the doping may have no independent effect and the change in $I_{\rm inel}$ is entirely due to the higher $R$ used on the lower doped sample.

In ref.~[\onlinecite{KhajetooriansNature2010}], a technique complementary to ISTS for probing the magnetic properties of an adsorbate on the InSb(110) surface was introduced. The asymmetry of the spin-split Landau level (LL) peaks of the Fe-induced 2DES was shown to reflect the adsorbate's magnetization as a function of magnetic field applied perpendicular to the surface. Fig.~\ref{fig:2} (a),(b) shows typical $\text{d}I/\text{d}V$ spectra, utilizing a W tip with negligible spin polarization, taken above the Fe adsorbate in a magnetic field for the higher and lower doped samples. Both spectra show pairs of peaks with intensities $\text{d}I/\text{d}V_{\uparrow}$ and $\text{d}I/\text{d}V_{\downarrow}$ in the entire voltage range above $E_{\rm 1}$ which are due to the spin-split LLs as marked by the corresponding arrows indicating the direction of the electron spin in each LL. Across the whole energy range, the spin-split peaks of each LL exhibit a large asymmetry $A_{\rm LL} = \frac{\text{d}I/\text{d}V_{\downarrow}-\text{d}I/\text{d}V_{\uparrow}}{\text{d}I/\text{d}V_{\downarrow}+\text{d}I/\text{d}V_{\uparrow}}$ (approx. $\unit[30]{\%}$ for the lowest LL at $B=\unit[5]{T}$), with the lower energy spin-split peak having the lower intensity for both samples. As shown in ref.~[\onlinecite{KhajetooriansNature2010}], this means that Fe majority spins are preferentially transmitted in the entire energy range from $\unit[-50]{meV}$ to $\unit[+100]{meV}$.


To understand the variations in the measured inelastic contribution to the ISTS spectra, we adopted the generalized version of the model used in ref.~[\onlinecite{PhysRevB.81.165423}], which we describe in the following.
The spin-dependent part of the Hamiltonian $\hat{H}$ for a magnetic adsorbate on a surface can be approximated by
$
\hat{H}=-g\mu_B\mathbf{B}\cdot\mathbf{\hat{S}}+D\hat{S}_z^2+E(\hat{S}_x^2-\hat{S}_y^2)
$
~\cite{Gatteschi}, where $\mathbf{\hat{S}}=(\hat{S}_x,\hat{S}_y,\hat{S}_z)$ is the vector containing the operators of the adsorbate spin, $\mathbf{B}$ is the magnetic field and $D$ and $E$ are the magnetic anisotropy energies resulting from the spin-orbit interaction of the adsorbate spin with its surroundings. In the experimental situation discussed here, the magnetic field always points along the $y$-axis (out of plane), i.e. $\mathbf{B}\cdot\hat{\mathbf{S}}=B_y \hat{S_y}$. Consequently, we chose this axis as the spin quantization axis. 

In ISTS, steps in the differential conductance appear due to the opening of additional tunneling channels when the tunneling electron's energy equals the energy difference between the magnetic ground state and an excited spin-state of the adsorbate. Consequently, $D$ and $E$ can be determined directly from the measured step energies. The step heights, on the other hand, depend on the excitation mechanism. Extracting the magnetic information encoded in the step heights thus requires a quantum mechanical model that links the underlying excitation mechanism to the experimentally observable quantities~\cite{Fernandez_Rossier_PhysRevLett.102.256802}.

In the model proposed in refs.~[\onlinecite{LorentePhysRevLett.103.176601,PhysRevB.81.165423}], the tunneling electron forms an intermediate total spin state with the magnetic adsorbate's spin $S$ during the excitation process. This state has spin $J=S\pm\frac{1}{2}$. The adsorbate states are elements of the spin-$S$ state space $\HH_S$ and the tunneling electron spin states are elements of the spin-$\frac{1}{2}$ state space $\HH_{1/2}$. Let $\{\phi_j \,|\, j=1,\dots,2S+1\}$ be an orthonormal basis of $\HH_S$, where the $\phi_j$ are eigenstates of $\hat{H}$. In the usual notation, we define $\langle \phi,m| = \langle\phi |\otimes\langle m|\in\HH_S \otimes \HH_{1/2}$ for $\langle\phi|\in\HH_S$ and $\langle m|\in\HH_{1/2}$.

The term
\begin{equation}
X_{i,m,f,m'}=|\langle\phi_i,m|\,c\,\OO_+ + (1-c)\,e^{i\varphi}\,\OO_-\,|\phi_f,m'\rangle|^2,
\label{eq:transitionIntensitysimple}
\end{equation}
with $c\in[0,1]$ and $\varphi\in[-\pi,+\pi]$, describes the relative transition intensity of a tunneling event in which a tunneling electron in the initial state $|m\rangle$ interacts with the magnetic adsorbate in the initial state $|\phi_i\rangle$. After the tunneling event, the adsorbate is in the final state $|\phi_f\rangle$ and the electron is in the final state $|m'\rangle$. These transitions are governed by the operators 
$
\OO_{\pm}=\sum_{M_J} |S\pm\nicefrac{1}{2},M_J\rangle\langle S\pm\nicefrac{1}{2},M_J|,
$
which describe the tunneling event as the formation of an intermediate state $|J,M_J\rangle$ with total spin $J=S\pm\nicefrac{1}{2}$ and magnetic quantum number $M_J$. The strength of the transition is proportional to the product of overlaps of the initial states with the intermediate state and the intermediate states with the final state. The summation over the different unobservable intermediate states is performed before taking the absolute square, so the different tunnel paths interfere. Note, that this model allows for a situation, where the electron transmission is governed by a mixture of majority ($J = S+1/2$) and minority ($J = S-1/2$) spin orbitals in the LDOS of the Fe adsorbate. The parameter $c$ governs the relative strength and $\varphi$ the relative phase of the contribution of these two spin channels. 

Before the model is analyzed in detail, we compare to other approaches reported in the literature. In refs.~[\onlinecite{LothNaturePhysics2010, NewJPhys.12.125021}], a similar model was used, where the relative transition intensities are given by the expression $|\langle \phi_i,m|\hat{\mathbf{S}}\cdot \hat{\mathbf{s}} + u\cdot\unity|\phi_f,m' \rangle|^2$ where $\hat{\mathbf{s}} = (\sigma_x,\sigma_y,\sigma_z)$ is the vector containing the electron spin Pauli matrices, $\unity$ is the identity operator. From the properties of $\OO_{\pm}$ given below, it is possible to derive the relation 
$
A_+\OO_+ + A_-\OO_- \propto  \hat{\mathbf{S}}\cdot \hat{\mathbf{s}} + \frac{A_+(S+1) +A_- S}{2(A_+-A_-)},
$
which shows that the transition operators defining the two models can be mapped onto each other, with the transformation from $(A_+,A_-)$ to $u$ given by the second term on the right hand side. 
 
We proceed by analyzing $I_{\rm inel}$ and $A_{\rm LL}$ as predicted by these models. At low temperature, only the ground state $|\phi_g\rangle$ will be occupied if we neglect artificial pumping of the spin states by the tunnel current as justified in ref.~[\onlinecite{KhajetooriansNature2010}]. 
If the ground state is non-degenerate, the elastic contribution to the transmission is proportional to $X_{\rm el}=\sum_m X_{g,m,g,m}$. This term describes all tunnel events in which the adsorbate's spin remains in the ground state during the collision, i.e. there is no excitation. Correspondingly, the transmission at energies higher than all excitations is proportional to $Z =\sum_{m} \sum_{f,m'} X_{g,m,f,m'}$, which describes all allowed events where the adsorbate starts in the ground state. The central quantity of interest in this work is the fraction of the differential conductance that is due to inelastic tunneling events, $I_{\rm inel}=1-\frac{X_{\rm el}}{Z}$. This expression can be generalized to the case of a degenerate ground state by summing over the different ground states in $X_{\rm el}$ and $Z$. For simplicity, we will assume a non-degenerate ground state in the following, as this is the case in our experimental situation~\cite{KhajetooriansNature2010}.   

Using the fact that the intermediate states $|J,M_J\rangle$ form a complete and orthonormal basis set  of $\mathcal{H}_S \otimes \mathcal{H}_{1/2}$, one can show that
$\OO_+ + \OO_- = \unity$, 
$\OO_+\OO_- = 0$,
$\OO_{\pm}^2 = \OO_{\pm}$, and
$\hat{\mathbf{S}}\cdot \hat{\mathbf{s}} =\frac{1}{2}S\OO_+ - \frac{1}{2}(S+1)\OO_-$. 
The expectation value of the spin scalar product $\Ssm{i}{m}=\langle\phi_i,m|\hat{\mathbf{S}}\cdot \hat{\mathbf{s}}|\phi_i,m\rangle$ is an important quantity in this model as will become clear below.
Because all terms in $\hat{H}$ except the Zeeman energy are quadratic in the adsorbate spin operators, it is zero at $\mathbf{B}=0$. It reaches its maximum magnitude $\max |\Ssm{i}{m}| = \nicefrac{1}{2} S$ when the eigenstates of $\hat{H}$ are simultaneous eigenstates of $\hat{S}_y$, and it is evaluated for an eigenstate with the magnetic quantum number $M=\pm S$.
It satisfies $\Ssm{i}{+\frac{1}{2}}=-\Ssm{i}{-\frac{1}{2}}$. 

Using the above relations, one can show that $I_{\rm inel}$ has the following properties for all $D,E,\mathbf{B}$: 
(i) phase symmetry $I_{\rm inel}(c,\varphi)=I_{\rm inel}(c,-\varphi)$. Therefore we limit all further investigations to the domain $\varphi\in[0,\pi]$. 
(ii) A global minimum $I_{\rm inel}=0$ occurs at $c=0.5, \varphi=0$, 
(iii) a global maximum $I_{\rm inel}=\frac{S^2 + S - 4\Ssm{g}{\nicefrac{1}{2}}^2}{S(S+1)}$ occurs at $c=\nicefrac{S}{2S+1}, \varphi=\pi$. For $\mathbf{B}=0$, this is maximal, $I_{\rm inel}=1$, and if the adsorbate's spin is fully aligned to the magnetic field, this maximum reaches its smallest possible value $\nicefrac{1}{S+1}$. (iv) There are no further extrema and $I_{\rm inel}$ increases with $\varphi$, except at the boundaries $c=0$ and $c=1$, where $I_{\rm inel}$ is independent of $\varphi$,
$
 I_{\rm inel}(c=0)=\frac{S^2+S-4\Ssm{g}{\frac{1}{2}}^2}{2S(S+\frac{1}{2})}
$
, and 
$
 I_{\rm inel}(c=1)=\frac{S^2+S-4\Ssm{g}{\frac{1}{2}}^2}{2(S+1)(S+\frac{1}{2})}
$.
These results imply that for $\mathbf{B}=0$ and at the maximum possible phase difference $\varphi=\pi$, if the intermediate spin channels contribute with just the right mixing $c=\nicefrac{S}{2S+1}$ between majority and minority spin orbitals, the elastic transmission is completely suppressed and the excitation efficiency is $\unit[100]{\%}$. On the other hand, for all magnetic fields, if the channels contribute equally with no phase difference, all excitation channels cancel out and only elastic tunnel events occur. 

\begin{figure}
\includegraphics[width=8.6cm]{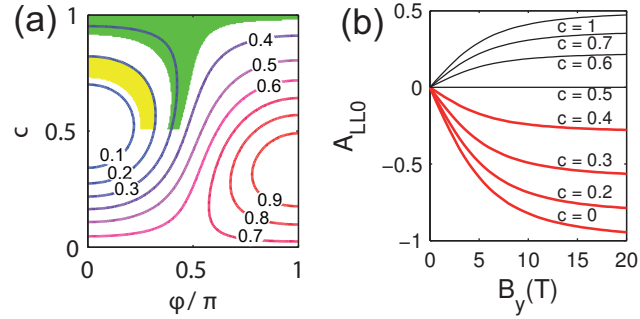}
\caption{
Model predictions for an adsorbate with spin $S=1$ and anisotropy energies $D=\unit[-1.4]{meV}$, $E=\unit[0.22]{meV}$.
(a) Contour plot of $I_{\rm inel}$ as a function of the model parameters $c$ and $\varphi$. Shaded areas highlight the experimentally relevant range of parameters for high/low doping in green/yellow
(b) Landau level asymmetry $A_{\rm LL}$  as a function of magnetic field $B_y$ along the $y$-axis for various values of the mixing parameter $c$. \label{fig:3}}
\end{figure}

The behavior of $I_{\rm inel}$ as a function of $c$ and $\varphi$ for the experimentally relevant S = 1 and the anisotropy energies that were found for Fe on InSb(110)~\cite{KhajetooriansNature2010} at $\mathbf{B}$=0 is shown in Fig.~\ref{fig:3}(a). 
The global maximum is located at $c=\nicefrac{1}{3}$, $\varphi=\pi$, and $I_{\rm inel}$ takes on all values in $[0,1]$. Note that every contour of constant $I_{\rm inel}$ intersects at least one of the boundaries $\varphi=0$, $\varphi=\pi$, so any value of $I_{\rm inel}$ can be produced by real coefficients of $\OO_{\pm}$. 

It is possible to show that the ratios of the inelastic steps
$\sum_{m,m'}X_{g,m,f_1,m'}/\sum_{m,m'}X_{g,m,f_2,m'}$ for $f_1,f_2 \neq g$
are independent of $c$ and $\varphi$, so these do not provide an independent source of information about the excitation mechanism. We therefore conclude that knowledge of $I_{\rm inel}$ at one magnetic field alone only implies which \textit{contour} on the $c$-$\varphi$-plane describes the tunnel conditions. To further narrow down the range of parameters, a complementary source of information is required. What is lacking is an observable that reflects the evolution of the adsorbate state in a magnetic field. 

The spin-split LLs shown in Fig.~\ref{fig:1}(a,b) provide a means of selecting the initial spin state of the tunneling electron, by setting the (negative) sample bias voltage to either the position of the spin-up or the spin-down peak. If the completely spin polarized tunnel current from one of these peaks is passed through a magnetic adsorbate on its way to the tip, it may interact with the local moment, creating spin excitations of the adsorbate and flipping the tunneling electron spin as a result. If the local moment is (partly) aligned by the magnetic field, this creates asymmetric tunnel conditions for the different tunnel electron spin orientations. As a result, the observed intensity of the two spin split LL peaks will be different. 
In ref.~[\onlinecite{KhajetooriansNature2010}], the model \eqref{eq:transitionIntensitysimple} was adapted to describe the LL asymmetry $A_{\rm LL}$for the case $S=1$. Using a similar analysis as for $I_{\rm inel}$, it is possible to derive a rather general expression for $A_{\rm LL}$.
Let $p_+$ be the relative peak height that corresponds to tunnel electrons that have spin up in the initial state and let $p_-$ correspond to spin down. These quantities are given by the total transition intensity for the adsorbate in the ground state and the electron in the up (down) state to be transmitted into any other configuration:
$
p_{\pm}=\sum_{f,m'}| \langle \phi_g,\pm\frac{1}{2}|c\OO_+ + e^{i\varphi}(1-c)\OO_-|\phi_f,m' \rangle |^2
$, since the energy of the lowest LL is much greater than all excitation energies, i.e. all channels are open. With this, we get the LL asymmetry
\[
A_{\rm LL}=\frac{p_- - p_+}{p_- + p_+} = \frac{-2(2c-1)}{c^2(2S+1)+S(1-2c)}\Ssm{g}{+\frac{1}{2}}.
\]
This is proportional to the projection of the adsorbate's spin on the magnetic field axis $\langle\phi_g|\hat{S_y}|\phi_g\rangle$, i.e. to the magnetization of the adsorbate. $A_{\rm LL}$ undergoes a sign change at $c=0.5$, so the sign of $A_{\rm LL}$ determines whether contours in the upper or lower half of Fig.~\ref{fig:3}(a) have to be considered.

Fig.~\ref{fig:3}(b) shows the predicted asymmetry curves for the experimentally relevant parameters of Fe adsorbates on InSb(110) \cite{KhajetooriansNature2010} and several values of $c$.
We note that in principle, if $A_{\rm LL}$ were known experimentally with high precision, $c$ could be determined precisely from Fig.~\ref{fig:3}(b) and then, using the the measured $I_{\rm inel}$ and Fig.~\ref{fig:3}(a), $\varphi$ would be known, too. This demonstrates the complementarity of $I_{\rm inel}$ and $A_{\rm LL}$. However, for $c$ close to either 1 or 0, the asymmetry curves are hard to distinguish experimentally, so it is unpractical to determine $c$ with high precision. From the positivity of our measured $A_{\rm LL}$, we deduce that $c>0.5$, i.e. the $S+\frac{1}{2}$ total spin channel dominates the tunneling. In other words, the majority spin orbitals dominate the Fe vacuum LDOS in the whole energy window.
From the measured $I_{\rm inel}$ for the higher/lower doped sample at $R=\unit[30]{M\Omega}$/$\unit[50]{M\Omega}$, we conclude that the parameters describing our system lie in the area highlighted in green/yellow in Fig.~\ref{fig:3}(a). This result shows how in principle the measured $I_{\rm inel}$ and $A_{\rm LL}$ can be used to extract the parameters $c$ and $\varphi$.

In the following, we discuss the results. The value of the parameters $c$ and $\varphi$ are determined by the details of the excitation process, namely whether tunneling electron and adsorbate spin preferentially couple parallel or antiparallel to form the intermediate total spin state and, if both channels are present, with which relative phase they contribute. Parallel coupling of the adsorbate and electron spins is facilitated if the vacuum LDOS above the adsorbate is dominated by majority spin orbitals and  antiparallel coupling is promoted by a minority dominated vacuum LDOS. In this way, $c$ reflects a crucial magnetic property of the adsorbate's electronic structure. Additionally, the formalism can be used to describe so called by-tunneling, where the tunnel electron does not couple to the adsorbate at all but \textit{bypasses} the magnetic adsorbate orbitals, leaving the adsorbate and electron in their initial spin states. This can be achieved by adding a multiple of the identity operator $\unity$:  $c\,\OO_+ + (1-c)\,e^{i\varphi}\,\OO_- + b\cdot\unity$. But since $\unity=\OO_+ + \OO_- $, this additional term can be absorbed into the coefficients of the $\OO_{\pm}$ and the shape of the transition operator remains unchanged, $a(c'\OO_+ + e^{i\varphi'}(1-c')\OO_-)$, where the scale factor $a$ has no effect on the relative transition intensities. This implies that the effect of by-tunneling on $I_{\rm inel}$ and $A_{\rm LL}$ cannot be distinguished from spin channel mixing. However, as we will discuss next, in our case, by-tunneling is better suited to explain the experimental observations.

In principle, the variation of band bending or junction resistance can influence $I_{\rm inel}$ in the following ways: (i) As illustrated in Fig.~\ref{fig:1}(c)/(d), the spin polarization of the vacuum LDOS at the Fermi energy may be energy dependent, resulting in changes in the mixing parameter $c$ as $E_{\rm F}$ varies relative to the adsorbate LDOS. Though an intriguing possibility for tuning the excitation efficiency, this is unlikely to be the case in our experiment, since the difference in band bending ($\Delta_{\rm BB}=\unit[25]{meV}$) is too small to produce a $\pi/4$ change in $\varphi$ or a  change of approx. $\unit[20]{\%}$ in the spin-polarization, which are the values that could account for the experimentally observed change in $I_{\rm inel}$ (see Fig.3(a)). (ii) Since the 2DES density is larger for the lower doping and tunneling into the substrate is presumably dominated by the 2DES, the tip-sample separation will be larger for lower doping. Likewise, increasing the junction resistance $R$ will increase the tip-sample separation $z$. If the vacuum spin polarization varies with distance from the surface, this may account for changing $c$. However, it seems unlikely that the vacuum spin polarization changes sufficiently on the distance scale of a few tens of $\unit{pm}$~\cite{PhysRevB.82.054411}, which is a rough estimate of the relevant change in $z$ by $R\propto\exp(2z/\unit{\AA})$. A more plausible explanation is that (iii) the by-tunneling contribution may be strongly distance dependent, because the Fe adsorption site lies below the topmost In and Sb layer~\cite{KhajetooriansNature2010}. Our experiment thus shows, that by-tunneling is getting stronger for larger tip-sample separation, which seems plausible.

Our method of the determination of $c$ and $\varphi$ that describe the excitation process will be most effective in a setting where by-tunneling is known to be minimal, so that the pure spin channel mixing can be observed. This is expected to be the case for adsorption geometries where the magnetic adsorbate is not strongly relaxed into the surface. 
We note that the measured $I_{\rm inel}=\unit[31]{\%}\pm\unit[4]{\%}$ is compatible with the value $I_{\rm inel}(c=1)=\frac{S}{2S+1}=\frac{1}{3}$ for $S=1$, $\mathbf{B}=0$. This on the one hand justifies the choice of $c=1$ in ref.~[\onlinecite{KhajetooriansNature2010}] and on the other hand shows that for low junction resistances on the higher doped sample, there is practically no by-tunneling.

In conclusion, we have shown how the inelastic contribution to ISTS spectra $I_{\rm inel}$ and the Landau level asymmetry $A_{\rm LL}$ can be used to investigate the mechanism of the spin excitation process. 
Our results motivate further investigations into the possibility of tuning the excitation efficiency. In particular, it would be desirable to obtain an independent measure of the by-tunneling with spatial and energetic resolution, as this cannot be distinguished from basic elastic tunneling through the magnetic adsorbate at this stage. 

We acknowledge funding from SFB668-A1 and
GrK1286 of the DFG, from the ERC Advanced Grant
FURORE, and from the Cluster of Excellence
Nanospintronics funded by the Forschungs- und
Wissenschaftsstiftung Hamburg.


\begin{thebibliography}{22}%
\makeatletter
\providecommand \@ifxundefined [1]{%
 \@ifx{#1\undefined}
}%
\providecommand \@ifnum [1]{%
 \ifnum #1\expandafter \@firstoftwo
 \else \expandafter \@secondoftwo
 \fi
}%
\providecommand \@ifx [1]{%
 \ifx #1\expandafter \@firstoftwo
 \else \expandafter \@secondoftwo
 \fi
}%
\providecommand \natexlab [1]{#1}%
\providecommand \enquote  [1]{``#1''}%
\providecommand \bibnamefont  [1]{#1}%
\providecommand \bibfnamefont [1]{#1}%
\providecommand \citenamefont [1]{#1}%
\providecommand \href@noop [0]{\@secondoftwo}%
\providecommand \href [0]{\begingroup \@sanitize@url \@href}%
\providecommand \@href[1]{\@@startlink{#1}\@@href}%
\providecommand \@@href[1]{\endgroup#1\@@endlink}%
\providecommand \@sanitize@url [0]{\catcode `\\12\catcode `\$12\catcode
  `\&12\catcode `\#12\catcode `\^12\catcode `\_12\catcode `\%12\relax}%
\providecommand \@@startlink[1]{}%
\providecommand \@@endlink[0]{}%
\providecommand \url  [0]{\begingroup\@sanitize@url \@url }%
\providecommand \@url [1]{\endgroup\@href {#1}{\urlprefix }}%
\providecommand \urlprefix  [0]{URL }%
\providecommand \Eprint [0]{\href }%
\providecommand \doibase [0]{http://dx.doi.org/}%
\providecommand \selectlanguage [0]{\@gobble}%
\providecommand \bibinfo  [0]{\@secondoftwo}%
\providecommand \bibfield  [0]{\@secondoftwo}%
\providecommand \translation [1]{[#1]}%
\providecommand \BibitemOpen [0]{}%
\providecommand \bibitemStop [0]{}%
\providecommand \bibitemNoStop [0]{.\EOS\space}%
\providecommand \EOS [0]{\spacefactor3000\relax}%
\providecommand \BibitemShut  [1]{\csname bibitem#1\endcsname}%
\let\auto@bib@innerbib\@empty
\bibitem [{\citenamefont {Heinrich}\ \emph {et~al.}(2004)\citenamefont
  {Heinrich}, \citenamefont {Gupta}, \citenamefont {Lutz},\ and\ \citenamefont
  {Eigler}}]{HeinrichScience2004}%
  \BibitemOpen
  \bibfield  {author} {\bibinfo {author} {\bibfnamefont {A.~J.}\ \bibnamefont
  {Heinrich}}, \bibinfo {author} {\bibfnamefont {J.~A.}\ \bibnamefont {Gupta}},
  \bibinfo {author} {\bibfnamefont {C.~P.}\ \bibnamefont {Lutz}}, \ and\
  \bibinfo {author} {\bibfnamefont {D.~M.}\ \bibnamefont {Eigler}},\ }\href
  {\doibase 10.1126/science.1101077} {\bibfield  {journal} {\bibinfo  {journal}
  {Science}\ }\textbf {\bibinfo {volume} {306}},\ \bibinfo {pages} {466}
  (\bibinfo {year} {2004})}\BibitemShut {NoStop}%
\bibitem [{\citenamefont {Hirjibehedin}\ \emph {et~al.}(2007)\citenamefont
  {Hirjibehedin}, \citenamefont {Lin}, \citenamefont {Otte}, \citenamefont
  {Ternes}, \citenamefont {Lutz}, \citenamefont {Jones},\ and\ \citenamefont
  {Heinrich}}]{HirjibehedinScience2007}%
  \BibitemOpen
  \bibfield  {author} {\bibinfo {author} {\bibfnamefont {C.~F.}\ \bibnamefont
  {Hirjibehedin}}, \bibinfo {author} {\bibfnamefont {C.-Y.}\ \bibnamefont
  {Lin}}, \bibinfo {author} {\bibfnamefont {A.~F.}\ \bibnamefont {Otte}},
  \bibinfo {author} {\bibfnamefont {M.}~\bibnamefont {Ternes}}, \bibinfo
  {author} {\bibfnamefont {C.~P.}\ \bibnamefont {Lutz}}, \bibinfo {author}
  {\bibfnamefont {B.~A.}\ \bibnamefont {Jones}}, \ and\ \bibinfo {author}
  {\bibfnamefont {A.~J.}\ \bibnamefont {Heinrich}},\ }\href {\doibase
  10.1126/science.1146110} {\bibfield  {journal} {\bibinfo  {journal}
  {Science}\ }\textbf {\bibinfo {volume} {317}},\ \bibinfo {pages} {1199}
  (\bibinfo {year} {2007})}\BibitemShut {NoStop}%
\bibitem [{\citenamefont {Chen}\ \emph {et~al.}(2008)\citenamefont {Chen},
  \citenamefont {Fu}, \citenamefont {Ji}, \citenamefont {Zhang}, \citenamefont
  {Cheng}, \citenamefont {Ma}, \citenamefont {Zou}, \citenamefont {Duan},
  \citenamefont {Jia},\ and\ \citenamefont {Xue}}]{PhysRevLett.101.197208}%
  \BibitemOpen
  \bibfield  {author} {\bibinfo {author} {\bibfnamefont {X.}~\bibnamefont
  {Chen}}, \bibinfo {author} {\bibfnamefont {Y.-S.}\ \bibnamefont {Fu}},
  \bibinfo {author} {\bibfnamefont {S.-H.}\ \bibnamefont {Ji}}, \bibinfo
  {author} {\bibfnamefont {T.}~\bibnamefont {Zhang}}, \bibinfo {author}
  {\bibfnamefont {P.}~\bibnamefont {Cheng}}, \bibinfo {author} {\bibfnamefont
  {X.-C.}\ \bibnamefont {Ma}}, \bibinfo {author} {\bibfnamefont {X.-L.}\
  \bibnamefont {Zou}}, \bibinfo {author} {\bibfnamefont {W.-H.}\ \bibnamefont
  {Duan}}, \bibinfo {author} {\bibfnamefont {J.-F.}\ \bibnamefont {Jia}}, \
  and\ \bibinfo {author} {\bibfnamefont {Q.-K.}\ \bibnamefont {Xue}},\ }\href
  {\doibase 10.1103/PhysRevLett.101.197208} {\bibfield  {journal} {\bibinfo
  {journal} {Phys. Rev. Lett.}\ }\textbf {\bibinfo {volume} {101}},\ \bibinfo
  {pages} {197208} (\bibinfo {year} {2008})}\BibitemShut {NoStop}%
\bibitem [{\citenamefont {Tsukahara}\ \emph {et~al.}(2009)\citenamefont
  {Tsukahara}, \citenamefont {Noto}, \citenamefont {Ohara}, \citenamefont
  {Shiraki}, \citenamefont {Takagi}, \citenamefont {Takata}, \citenamefont
  {Miyawaki}, \citenamefont {Taguchi}, \citenamefont {Chainani}, \citenamefont
  {Shin},\ and\ \citenamefont {Kawai}}]{KawaiPRL2010}%
  \BibitemOpen
  \bibfield  {author} {\bibinfo {author} {\bibfnamefont {N.}~\bibnamefont
  {Tsukahara}}, \bibinfo {author} {\bibfnamefont {K.-i.}\ \bibnamefont {Noto}},
  \bibinfo {author} {\bibfnamefont {M.}~\bibnamefont {Ohara}}, \bibinfo
  {author} {\bibfnamefont {S.}~\bibnamefont {Shiraki}}, \bibinfo {author}
  {\bibfnamefont {N.}~\bibnamefont {Takagi}}, \bibinfo {author} {\bibfnamefont
  {Y.}~\bibnamefont {Takata}}, \bibinfo {author} {\bibfnamefont
  {J.}~\bibnamefont {Miyawaki}}, \bibinfo {author} {\bibfnamefont
  {M.}~\bibnamefont {Taguchi}}, \bibinfo {author} {\bibfnamefont
  {A.}~\bibnamefont {Chainani}}, \bibinfo {author} {\bibfnamefont
  {S.}~\bibnamefont {Shin}}, \ and\ \bibinfo {author} {\bibfnamefont
  {M.}~\bibnamefont {Kawai}},\ }\href {\doibase 10.1103/PhysRevLett.102.167203}
  {\bibfield  {journal} {\bibinfo  {journal} {Phys. Rev. Lett.}\ }\textbf
  {\bibinfo {volume} {102}},\ \bibinfo {pages} {167203} (\bibinfo {year}
  {2009})}\BibitemShut {NoStop}%
\bibitem [{\citenamefont {Meier}\ \emph {et~al.}(2008)\citenamefont {Meier},
  \citenamefont {Zhou}, \citenamefont {Wiebe},\ and\ \citenamefont
  {Wiesendanger}}]{MZW2008}%
  \BibitemOpen
  \bibfield  {author} {\bibinfo {author} {\bibfnamefont {F.}~\bibnamefont
  {Meier}}, \bibinfo {author} {\bibfnamefont {L.}~\bibnamefont {Zhou}},
  \bibinfo {author} {\bibfnamefont {J.}~\bibnamefont {Wiebe}}, \ and\ \bibinfo
  {author} {\bibfnamefont {R.}~\bibnamefont {Wiesendanger}},\ }\href@noop {}
  {\bibfield  {journal} {\bibinfo  {journal} {Science}\ }\textbf {\bibinfo
  {volume} {320}},\ \bibinfo {pages} {82} (\bibinfo {year} {2008})}\BibitemShut
  {NoStop}%
\bibitem [{\citenamefont {Hirjibehedin}\ \emph {et~al.}(2006)\citenamefont
  {Hirjibehedin}, \citenamefont {Lutz},\ and\ \citenamefont
  {Heinrich}}]{HirjibehedinScience2006}%
  \BibitemOpen
  \bibfield  {author} {\bibinfo {author} {\bibfnamefont {C.~F.}\ \bibnamefont
  {Hirjibehedin}}, \bibinfo {author} {\bibfnamefont {C.~P.}\ \bibnamefont
  {Lutz}}, \ and\ \bibinfo {author} {\bibfnamefont {A.~J.}\ \bibnamefont
  {Heinrich}},\ }\href {\doibase 10.1126/science.1125398} {\bibfield  {journal}
  {\bibinfo  {journal} {Science}\ }\textbf {\bibinfo {volume} {312}},\ \bibinfo
  {pages} {1021} (\bibinfo {year} {2006})}\BibitemShut {NoStop}%
\bibitem [{\citenamefont {Zhou}\ \emph {et~al.}(2010)\citenamefont {Zhou},
  \citenamefont {Wiebe}, \citenamefont {Lounis}, \citenamefont {Vedmedenko},
  \citenamefont {Meier}, \citenamefont {Bl{\"u}gel}, \citenamefont
  {Dederichs},\ and\ \citenamefont {Wiesendanger}}]{ZhouWiebeLounis2010}%
  \BibitemOpen
  \bibfield  {author} {\bibinfo {author} {\bibfnamefont {L.}~\bibnamefont
  {Zhou}}, \bibinfo {author} {\bibfnamefont {J.}~\bibnamefont {Wiebe}},
  \bibinfo {author} {\bibfnamefont {S.}~\bibnamefont {Lounis}}, \bibinfo
  {author} {\bibfnamefont {E.}~\bibnamefont {Vedmedenko}}, \bibinfo {author}
  {\bibfnamefont {F.}~\bibnamefont {Meier}}, \bibinfo {author} {\bibfnamefont
  {S.}~\bibnamefont {Bl{\"u}gel}}, \bibinfo {author} {\bibfnamefont {P.~H.}\
  \bibnamefont {Dederichs}}, \ and\ \bibinfo {author} {\bibfnamefont
  {R.}~\bibnamefont {Wiesendanger}},\ }\href@noop {} {\bibfield  {journal}
  {\bibinfo  {journal} {Nat. Phys.}\ }\textbf {\bibinfo {volume} {6}},\
  \bibinfo {pages} {187} (\bibinfo {year} {2010})}\BibitemShut {NoStop}%
\bibitem [{\citenamefont {Lorente}\ and\ \citenamefont
  {Gauyacq}(2009)}]{LorentePhysRevLett.103.176601}%
  \BibitemOpen
  \bibfield  {author} {\bibinfo {author} {\bibfnamefont {N.}~\bibnamefont
  {Lorente}}\ and\ \bibinfo {author} {\bibfnamefont {J.-P.}\ \bibnamefont
  {Gauyacq}},\ }\href {\doibase 10.1103/PhysRevLett.103.176601} {\bibfield
  {journal} {\bibinfo  {journal} {Phys. Rev. Lett.}\ }\textbf {\bibinfo
  {volume} {103}},\ \bibinfo {pages} {176601} (\bibinfo {year}
  {2009})}\BibitemShut {NoStop}%
\bibitem [{\citenamefont {Gauyacq}\ \emph {et~al.}(2010)\citenamefont
  {Gauyacq}, \citenamefont {Novaes},\ and\ \citenamefont
  {Lorente}}]{PhysRevB.81.165423}%
  \BibitemOpen
  \bibfield  {author} {\bibinfo {author} {\bibfnamefont {J.-P.}\ \bibnamefont
  {Gauyacq}}, \bibinfo {author} {\bibfnamefont {F.~D.}\ \bibnamefont {Novaes}},
  \ and\ \bibinfo {author} {\bibfnamefont {N.}~\bibnamefont {Lorente}},\ }\href
  {\doibase 10.1103/PhysRevB.81.165423} {\bibfield  {journal} {\bibinfo
  {journal} {Phys. Rev. B}\ }\textbf {\bibinfo {volume} {81}},\ \bibinfo
  {pages} {165423} (\bibinfo {year} {2010})}\BibitemShut {NoStop}%
\bibitem [{\citenamefont
  {Fern\'andez-Rossier}(2009)}]{Fernandez_Rossier_PhysRevLett.102.256802}%
  \BibitemOpen
  \bibfield  {author} {\bibinfo {author} {\bibfnamefont {J.}~\bibnamefont
  {Fern\'andez-Rossier}},\ }\href {\doibase 10.1103/PhysRevLett.102.256802}
  {\bibfield  {journal} {\bibinfo  {journal} {Phys. Rev. Lett.}\ }\textbf
  {\bibinfo {volume} {102}},\ \bibinfo {pages} {256802} (\bibinfo {year}
  {2009})}\BibitemShut {NoStop}%
\bibitem [{\citenamefont {Fransson}(2009)}]{Fransson_Nanolett2009}%
  \BibitemOpen
  \bibfield  {author} {\bibinfo {author} {\bibfnamefont {J.}~\bibnamefont
  {Fransson}},\ }\href@noop {} {\bibfield  {journal} {\bibinfo  {journal} {Nano
  Lett.}\ }\textbf {\bibinfo {volume} {9}},\ \bibinfo {pages} {2414} (\bibinfo
  {year} {2009})}\BibitemShut {NoStop}%
\bibitem [{\citenamefont {Persson}(2009)}]{Persson_PhysRevLett.103.050801}%
  \BibitemOpen
  \bibfield  {author} {\bibinfo {author} {\bibfnamefont {M.}~\bibnamefont
  {Persson}},\ }\href {\doibase 10.1103/PhysRevLett.103.050801} {\bibfield
  {journal} {\bibinfo  {journal} {Phys. Rev. Lett.}\ }\textbf {\bibinfo
  {volume} {103}},\ \bibinfo {pages} {050801} (\bibinfo {year}
  {2009})}\BibitemShut {NoStop}%
\bibitem [{\citenamefont {Loth}\ \emph
  {et~al.}(2010{\natexlab{a}})\citenamefont {Loth}, \citenamefont {Lutz},\ and\
  \citenamefont {Heinrich}}]{NewJPhys.12.125021}%
  \BibitemOpen
  \bibfield  {author} {\bibinfo {author} {\bibfnamefont {S.}~\bibnamefont
  {Loth}}, \bibinfo {author} {\bibfnamefont {C.~P.}\ \bibnamefont {Lutz}}, \
  and\ \bibinfo {author} {\bibfnamefont {A.~J.}\ \bibnamefont {Heinrich}},\
  }\href@noop {} {\bibfield  {journal} {\bibinfo  {journal} {New J. Phys.}\
  }\textbf {\bibinfo {volume} {12}},\ \bibinfo {pages} {125021} (\bibinfo
  {year} {2010}{\natexlab{a}})}\BibitemShut {NoStop}%
\bibitem [{\citenamefont {Schrieffer}\ and\ \citenamefont
  {Wolff}(1966)}]{PhysRev.149.491}%
  \BibitemOpen
  \bibfield  {author} {\bibinfo {author} {\bibfnamefont {J.~R.}\ \bibnamefont
  {Schrieffer}}\ and\ \bibinfo {author} {\bibfnamefont {P.~A.}\ \bibnamefont
  {Wolff}},\ }\href {\doibase 10.1103/PhysRev.149.491} {\bibfield  {journal}
  {\bibinfo  {journal} {Phys. Rev.}\ }\textbf {\bibinfo {volume} {149}},\
  \bibinfo {pages} {491} (\bibinfo {year} {1966})}\BibitemShut {NoStop}%
\bibitem [{\citenamefont {Appelbaum}(1967)}]{PhysRev.154.633}%
  \BibitemOpen
  \bibfield  {author} {\bibinfo {author} {\bibfnamefont {J.~A.}\ \bibnamefont
  {Appelbaum}},\ }\href {\doibase 10.1103/PhysRev.154.633} {\bibfield
  {journal} {\bibinfo  {journal} {Phys. Rev.}\ }\textbf {\bibinfo {volume}
  {154}},\ \bibinfo {pages} {633} (\bibinfo {year} {1967})}\BibitemShut
  {NoStop}%
\bibitem [{\citenamefont {De~Franceschi}\ \emph {et~al.}(2001)\citenamefont
  {De~Franceschi}, \citenamefont {Sasaki}, \citenamefont {Elzerman},
  \citenamefont {van~der Wiel}, \citenamefont {Tarucha},\ and\ \citenamefont
  {Kouwenhoven}}]{PhysRevLett.86.878}%
  \BibitemOpen
  \bibfield  {author} {\bibinfo {author} {\bibfnamefont {S.}~\bibnamefont
  {De~Franceschi}}, \bibinfo {author} {\bibfnamefont {S.}~\bibnamefont
  {Sasaki}}, \bibinfo {author} {\bibfnamefont {J.~M.}\ \bibnamefont
  {Elzerman}}, \bibinfo {author} {\bibfnamefont {W.~G.}\ \bibnamefont {van~der
  Wiel}}, \bibinfo {author} {\bibfnamefont {S.}~\bibnamefont {Tarucha}}, \ and\
  \bibinfo {author} {\bibfnamefont {L.~P.}\ \bibnamefont {Kouwenhoven}},\
  }\href {\doibase 10.1103/PhysRevLett.86.878} {\bibfield  {journal} {\bibinfo
  {journal} {Phys. Rev. Lett.}\ }\textbf {\bibinfo {volume} {86}},\ \bibinfo
  {pages} {878} (\bibinfo {year} {2001})}\BibitemShut {NoStop}%
\bibitem [{\citenamefont {Khajetoorians}\ \emph {et~al.}(2010)\citenamefont
  {Khajetoorians}, \citenamefont {Chilian}, \citenamefont {Wiebe},
  \citenamefont {Sergej~Schuwalow},\ and\ \citenamefont
  {Wiesendanger}}]{KhajetooriansNature2010}%
  \BibitemOpen
  \bibfield  {author} {\bibinfo {author} {\bibfnamefont {A.~A.}\ \bibnamefont
  {Khajetoorians}}, \bibinfo {author} {\bibfnamefont {B.}~\bibnamefont
  {Chilian}}, \bibinfo {author} {\bibfnamefont {J.}~\bibnamefont {Wiebe}},
  \bibinfo {author} {\bibfnamefont {F.~L.}\ \bibnamefont {Sergej~Schuwalow}}, \
  and\ \bibinfo {author} {\bibfnamefont {R.}~\bibnamefont {Wiesendanger}},\
  }\href@noop {} {\bibfield  {journal} {\bibinfo  {journal} {Nature}\ }\textbf
  {\bibinfo {volume} {467}},\ \bibinfo {pages} {1084} (\bibinfo {year}
  {2010})}\BibitemShut {NoStop}%
\bibitem [{\citenamefont {Wiebe}\ \emph {et~al.}(2004)\citenamefont {Wiebe},
  \citenamefont {Wachowiak}, \citenamefont {Meier}, \citenamefont {Haude},
  \citenamefont {Foster}, \citenamefont {Morgenstern},\ and\ \citenamefont
  {Wiesendanger}}]{WWM2004}%
  \BibitemOpen
  \bibfield  {author} {\bibinfo {author} {\bibfnamefont {J.}~\bibnamefont
  {Wiebe}}, \bibinfo {author} {\bibfnamefont {A.}~\bibnamefont {Wachowiak}},
  \bibinfo {author} {\bibfnamefont {F.}~\bibnamefont {Meier}}, \bibinfo
  {author} {\bibfnamefont {D.}~\bibnamefont {Haude}}, \bibinfo {author}
  {\bibfnamefont {T.}~\bibnamefont {Foster}}, \bibinfo {author} {\bibfnamefont
  {M.}~\bibnamefont {Morgenstern}}, \ and\ \bibinfo {author} {\bibfnamefont
  {R.}~\bibnamefont {Wiesendanger}},\ }\href@noop {} {\bibfield  {journal}
  {\bibinfo  {journal} {Rev. Sci. Instrum.}\ }\textbf {\bibinfo {volume}
  {75}},\ \bibinfo {pages} {4871} (\bibinfo {year} {2004})}\BibitemShut
  {NoStop}%
\bibitem [{\citenamefont {Hashimoto}\ \emph {et~al.}(2008)\citenamefont
  {Hashimoto}, \citenamefont {Sohrmann}, \citenamefont {Wiebe}, \citenamefont
  {Inaoka}, \citenamefont {Meier}, \citenamefont {Hirayama}, \citenamefont
  {R{\"o}mer}, \citenamefont {Wiesendanger},\ and\ \citenamefont
  {Morgenstern}}]{HSW2008}%
  \BibitemOpen
  \bibfield  {author} {\bibinfo {author} {\bibfnamefont {K.}~\bibnamefont
  {Hashimoto}}, \bibinfo {author} {\bibfnamefont {C.}~\bibnamefont {Sohrmann}},
  \bibinfo {author} {\bibfnamefont {J.}~\bibnamefont {Wiebe}}, \bibinfo
  {author} {\bibfnamefont {T.}~\bibnamefont {Inaoka}}, \bibinfo {author}
  {\bibfnamefont {F.}~\bibnamefont {Meier}}, \bibinfo {author} {\bibfnamefont
  {Y.}~\bibnamefont {Hirayama}}, \bibinfo {author} {\bibfnamefont {R.~A.}\
  \bibnamefont {R{\"o}mer}}, \bibinfo {author} {\bibfnamefont {R.}~\bibnamefont
  {Wiesendanger}}, \ and\ \bibinfo {author} {\bibfnamefont {M.}~\bibnamefont
  {Morgenstern}},\ }\href@noop {} {\bibfield  {journal} {\bibinfo  {journal}
  {Phys. Rev. Lett.}\ }\textbf {\bibinfo {volume} {101}},\ \bibinfo {pages}
  {256802} (\bibinfo {year} {2008})}\BibitemShut {NoStop}%
\bibitem [{\citenamefont {Gatteschi}\ \emph {et~al.}(2006)\citenamefont
  {Gatteschi}, \citenamefont {Sessoli},\ and\ \citenamefont
  {Villain}}]{Gatteschi}%
  \BibitemOpen
  \bibfield  {author} {\bibinfo {author} {\bibfnamefont {D.}~\bibnamefont
  {Gatteschi}}, \bibinfo {author} {\bibfnamefont {R.}~\bibnamefont {Sessoli}},
  \ and\ \bibinfo {author} {\bibfnamefont {J.}~\bibnamefont {Villain}},\
  }\href@noop {} {\emph {\bibinfo {title} {Molecular Nanomagnets}}}\ (\bibinfo
  {publisher} {Oxford Univ. Press, Oxford},\ \bibinfo {year}
  {2006})\BibitemShut {NoStop}%
\bibitem [{\citenamefont {Loth}\ \emph
  {et~al.}(2010{\natexlab{b}})\citenamefont {Loth}, \citenamefont {von
  Bergmann}, \citenamefont {Ternes}, \citenamefont {Otte}, \citenamefont
  {Lutz},\ and\ \citenamefont {Heinrich}}]{LothNaturePhysics2010}%
  \BibitemOpen
  \bibfield  {author} {\bibinfo {author} {\bibfnamefont {S.}~\bibnamefont
  {Loth}}, \bibinfo {author} {\bibfnamefont {K.}~\bibnamefont {von Bergmann}},
  \bibinfo {author} {\bibfnamefont {M.}~\bibnamefont {Ternes}}, \bibinfo
  {author} {\bibfnamefont {A.~F.}\ \bibnamefont {Otte}}, \bibinfo {author}
  {\bibfnamefont {C.~P.}\ \bibnamefont {Lutz}}, \ and\ \bibinfo {author}
  {\bibfnamefont {A.~J.}\ \bibnamefont {Heinrich}},\ }\href@noop {} {\bibfield
  {journal} {\bibinfo  {journal} {Nature Physics}\ }\textbf {\bibinfo {volume}
  {6}},\ \bibinfo {pages} {340} (\bibinfo {year}
  {2010}{\natexlab{b}})}\BibitemShut {NoStop}%
\bibitem [{\citenamefont {Ferriani}\ \emph {et~al.}(2010)\citenamefont
  {Ferriani}, \citenamefont {Lazo},\ and\ \citenamefont
  {Heinze}}]{PhysRevB.82.054411}%
  \BibitemOpen
  \bibfield  {author} {\bibinfo {author} {\bibfnamefont {P.}~\bibnamefont
  {Ferriani}}, \bibinfo {author} {\bibfnamefont {C.}~\bibnamefont {Lazo}}, \
  and\ \bibinfo {author} {\bibfnamefont {S.}~\bibnamefont {Heinze}},\ }\href
  {\doibase 10.1103/PhysRevB.82.054411} {\bibfield  {journal} {\bibinfo
  {journal} {Phys. Rev. B}\ }\textbf {\bibinfo {volume} {82}},\ \bibinfo
  {pages} {054411} (\bibinfo {year} {2010})}\BibitemShut {NoStop}%
\end{thebibliography}

%


\end{document}